# Inconsistencies of Recently Proposed Citation Impact Indicators and how to Avoid Them

Michael SCHREIBER

*Institute of Physics, Chemnitz University of Technology, 09107 Chemnitz, Germany.*
*E-mail: schreiber@physik.tu-chemnitz.de*

It is shown that under certain circumstances in particular for small datasets the recently proposed citation impact indicators $I3(6PR)$ and $R(6,k)$ behave inconsistently when additional papers or citations are taken into consideration. Three simple examples are presented, in which the indicators fluctuate strongly and the ranking of scientists in the evaluated group is sometimes completely mixed up by minor changes in the data base. The erratic behavior is traced to the specific way in which weights are attributed to the six percentile rank classes, specifically for the tied papers. For 100 percentile rank classes the effects will be less serious. For the 6 classes it is demonstrated that a different way of assigning weights avoids these problems, although the non-linearity of the weights for the different percentile rank classes can still lead to (much less frequent) changes in the ranking. This behavior is not undesired, because it can be used to correct for differences in citation behavior in different fields. Remaining deviations from the theoretical value $R(6,k) = 1.91$ can be avoided by a new scoring rule, the fractional scoring. Previously proposed consistency criteria are amended by another property of strict independence which a performance indicator should aim at.

**Introduction**

There has recently been a controversy about the best way of normalization of citation impact indicators. The intensive discussion in the literature in the last two years shall not be repeated here. A good overview is given by Leydesdorff, Bornmann, Mutz, and Opthof (2011). The core issue of that debate was about replacing the rate of averages with the average of rates. Finally two of the main contributors joined forces and proposed the use of alternative indicators based on non-parametric scores. They applied it to a set of 7 individuals (Leydesdorff et al., 2011) as well as to two larger sets of journals (Leydesdorff & Bornmann, 2011). But also the latter version is meant to enable one to compare citation impact not only among journals, but also among institutions or individuals.

Rousseau (2012) has already noted "that the I3 is not a consistent … indicator. This is because adding a new article changes the reference set, and hence, each element's percentile and all class borders. It is then always possible – by exploiting these small changes – to prove inconsistency" (p. 419). When I tried to apply the indicators to a rather small number of scientists with not so large citation records, specifically the 26 physicists whose citation records I have previously studied in detail (Schreiber, 2008, 2009) I observed such inconsistencies: small changes in the data base lead to unexpected behavior of the indicators. In order to illustrate these problems, I have constructed a few simple examples which are presented below. Thus I can show that additional papers and/or citations can lead to strongly fluctuating values of the indicators and can upset the ranking of scientists in the evaluated group, even if the citation records of these scientists remain unchanged. In several instances, these changes violate a consistency criterium which was recently proposed by Waltman and van Eck (2012), but in a variety of instances from my examples the counterintuitive behavior which I call inconsistent, is not covered by the criteria in that paper. Rather a new criterium is needed, as proposed in the final section of the present manuscript.

## Calculation of the Citation Impact Indicators *I*3(6PR) and *R*(6,*k*)

The new indicators are based on the determination of percentile ranks which are determined for each publication. Specifically, Leydesdorff and Bornmann (2011) propose the "counting rule that the number of items with lower citation rates than the item under study determines the percentile" (p. 2137). In Table 1 this is applied to my first example dataset A which comprises 40 publications. For the present purpose it is not yet necessary to distribute these publications among several individuals. This can be done later, at present the table remains easier to survey, if only the total dataset is described and all papers with the same number of citations are collected in one group. Therefore in Table 1 the number $n(c)$ of publications with a certain citation frequency $c$ is given. Altogether there are $n_{tot} = \sum_{c=0}^{\infty} n(c) = 40$ papers with altogether $n_{tot} = \sum_{c=0}^{\infty} c*n(c) = 52$ citations. For every group the number $n_<$ of papers with less citations than each paper in this group received, is determined. This is then expressed as a percentage. In accordance with Leydesdorff et al. (2011) "tied, that is, precisely equal, numbers of citations thus are not counted as fewer than" (p. 1372). The rounded integer numbers of the percentiles thus apply to all papers in the respective group, because they have exactly equal citation frequencies.

Following the categorization used by the National Science Board (2010) for *Science and Engineering Indicators* (Appendix table 5-43) Bornmann and Mutz (2011) suggested to categorize the papers into six percentile rank (PR) classes, namely the best 1%, 5%, 10%, 25%, 50%, and the remaining 50%.[1] To be specific, papers with a thus determined percentile less than the 50$^{th}$ (or, equivalently a percentage less than 50.0) are included in the class 1, and papers with a percentile less than the 75$^{th}$, 90$^{th}$, 95$^{th}$, 99$^{th}$ percentile are aggregated in class 2, 3, 4, 5, respectively, if they do not belong to a lower rank class. The top class 6 comprises the 99$^{th}$ and 100$^{th}$ percentile. These label numbers $w$ of the six PR classes are then employed to weight each publication as shown in Table 1. Using this scoring rule the sum over all publications was taken by Leydesdorff and Bornmann (2011) to define the integrated impact indicator $I3$(6PR). These authors have also used the percentiles directly as weights between 1 and 100 to define $I3$(100PR). This shall not be done in the present investigation. For the example case A1 in Table 1 one obtains $I3$(6PR) = $n_{tot} = \sum_{c=0}^{\infty} w*n(c) = 76$ by adding the values in the last line in the table. Normalizing the result by the total number of papers yields the citation impact indicator $R(6,k)$ when the same six PR classes are used (Leydesdorff et al., 2011). Thus $R(6,k) = 76/40 = 1.9$ for the example in Table 1.

**Inconsistencies of the Indicators for Example A**

Due to the relatively small number of source items in the example, and due to the relatively small number of citations, a large number of ties occurs. Below it is shown that the specific way of treating the tied papers can lead to counterintuitive and inconsistent behavior of $I3$(6PR) (and thus also of $R(6,k)$), which is aggravated by the non-linear behavior due to the use of 6 PR classes. If 100 PR classes are used, one can in principle construct similar examples, but the occurrence of inconsistencies is much more unlikely and the effects are smaller.

---

[1] In this case it appears to be more appropriate not to round the percentages to integer percentiles, because such a rounding would for example put the 49.5% into the 50$^{th}$ percentile and thus into the second lowest PR class instead of the lowest PR class. Therefore one should rather take the rational number of the percentage or "round" non-integer values to the next lower integer number (i.e. use the floor function) to define the percentile. The discussed rounding procedure does not have an influence on the values in Table 1.

The inconsistencies occur for the present example A already, when one assumes that one of the uncited papers in the example receives more and more citations, while the citation rates of the other papers remain unchanged.

If one of the uncited papers receives its first citation, then the resulting indicator $I3(6PR)$ significantly drops, namely to the value 66. The derivation of this result is shown in Table 2 in analogy to Table 1.[2] Assuming that this publication receives additional citations the $I3(6PR)$ indicator shows a fluctuating behavior, as demonstrated in Table 3. As the number of papers remains constant, the normalized indicator $R(6,k)$ behaves in the same way. Overall, a strong decrease of these indicators is achieved, although the impact of the dataset as measured by the number of citations is increasing. I consider this behavior counterintuitive and in my view therefore the indicator $I3(6PR)$ as well as $R(6,k)$ are inconsistent.

A closer inspection of Tables 1 and 2 reveals the reason for the strong drop of the indicators: For a large number of papers (namely those with one citation each) the percentage $n_<$ changes from 50.0 to 47.5 and thus their PR class changes so that their weight decreases from two to one. This is due to the specific treatment of tied papers, which are all grouped into the lower PR class, because there are not enough papers with fewer citations. This can be seen in Table 2 where the 11 papers with one citation each are all grouped into the $47^{th}$ percentile and thus into the bottom-50% class, although together with the 19 uncited papers they amount to 75% of the entire sample.

Leydesdorff and Bornmann (2011) stated that they "wish to give the papers the benefit of the doubt" (p. 2137), by providing tied papers with the highest possible ranks. I note that this is in contradiction to their stated strict counting rule. Overruling the counting rule and giving these 11 papers the highest possible rank by attributing them to the $72^{nd}$ percentile would give each of them a weight of 2 and result in a value of 77 for the indicator $I3(6PR)$. Thus this would make the behavior of $I3(6PR)$ look more consistent comparing Tables 1 and 2, because an additional citation would lead to only a small increase of $I3(6PR)$. (Ideally, I3(6PR) should not change at all for a different number of citations.) However, if one changes the citation numbers in Table 1 in the opposite way, namely to 21 uncited papers and 9 papers with one citation each, then all the 21 papers would end up in the $50^{th}$ percentile due to "the benefit of the doubt" and thus they would all get a weight of 2, resulting in a total $I3(6PR) = 96$. So the inconsistency would not only remain, but it would be enhanced.

---

[2] Here the usual rounding of the percentage 47.5 would put the singly cited papers into the $48^{th}$ percentile, while cutting by the floor function would put them into the $47^{th}$ percentile, without any change in the rank class and thus without any change for $I3(6PR)$.

Pudovkin and Garfield (2009) have used a different scheme to treat tied papers, namely assigning the average of the ranks to every paper in a tied set. Aside from rounding effects, this is equivalent to averaging the percentile values and does also not solve the underlying problem, because again the assignment of a large number of tied papers to one or the other PR class can be determined by a single (additional) citation. Changing the present example by starting with 12 uncited papers and 18 papers with one citation each, the respective percentile classes for the second group would range from 30 to 72 and give an average value of 51. This would put these papers into rank class 2 and attribute a weight of 2 to each of them, yielding a contribution of 36 to $I3(6PR)$ in addition to 12 points from the 12 uncited papers. A single additional citation to a previously uncited paper would put the now 19 papers with one citation each into the percentile classes from 27 to 72 with an average of 49.7 and therefore give each of these a weight of 1, resulting in a contribution of only 19 to $I3(6PR)$ in addition to 11 points from the 11 remaining uncited papers. Again an additional citation has led to a strong reduction of $I3(6PR)$.

**How to Avoid the Inconsistencies by Attributing Averaged Weights**

A solution of the present problem is not to take the lowest possible rank as in Tables 1 and 2 nor to take the highest possible rank as suggested by Leydesdorff and Bornmann (2011), neither to average the rank as proposed by Pudovkin and Garfield (2009) but rather to average the weights of the tied papers. This means to initially use different ranks for the papers (without caring about the sequence of tied papers) and to attribute respective weights according to the utilized scoring rule; then the arithmetic average is taken over the assigned weights for the tied papers and this average weight is now given to all the tied papers which will thus get the same value so that their sequence does indeed not matter. This means that non-integer weights are utilized and thus the above demonstrated strong jumps in the contribution of tied papers to the overall value of the indicator which I now label $I3^{av}(6PR)$ (and therefore also to the normalized $R^{av}(6,k)$) are avoided. In the present example we obtain $I3^{av}(6PR) = 76$ and $R^{av}(6,k) = 1.9$ for all cases in Table 3. It may be considered a disadvantage that the indicators do not change, although the number of citation changes. But this is not surprising, because the indicators are based on the ranking of the papers rather than the actual number of citations. Therefore, besides artificial effects due to the specific way of grouping and weighting tied papers and besides rounding effects due to small numbers in the finite dataset, the indicators for the complete dataset should not depend on the number of citations. Due to the normalization, $R^{av}(6,k)$ should not even depend on the number of papers. Of course, for different *sub*sets representing individual scientists there will be changes as demonstrated below.

**Another Example with Increasing Number of Publications**

If one does not keep the total number of papers in the sample constant, but allows additional uncited papers to enter the investigation, the inconsistencies of the indicators can be even stronger. This is visualized in Figure 1 where I present a second example B, starting with case B1 which comprises 7 singly cited papers, 4 doubly cited and 4 triply cited papers, i.e. altogether 15 publications with altogether 27 citations. The histogram in Figure 1 illustrates the number of papers for each citation frequency.

Adding another paper without any citation makes the indicator $I3(6PR)$ jump from 19 to 28, and likewise $R(6,k)$ from 1.27 to 1.75. Adding further uncited papers leads to an expected steady, but rather small increase of $I3(6PR)$. This is shown for the cases B2 to B15 in Figure 1. At the same time, $R(6,k)$ decreases, because of the normalization by means of the total number of papers.

The next additional uncited paper leads to a jump of $I3(6PR)$ from 85 to 98, because suddenly all singly cited papers have reached the second PR class. This is the case B16 in Figure 1. The jump can be compensated by attributing a first citation to one of the uncited papers, see case B17 in Figure 1.

In the following cases I have either added a single uncited paper to the example, or a single citation. In the latter situation the top of the histogram in Figure 1 remains flat, but one (and only one) of the internal boundaries in the histogram displays a step. The resulting behavior of $I3(6PR)$ in Figure 1 appears rather erratic. Admittedly, I have selected the specific increases in such a way as to maximize the fluctuations which can be seen in Figure 1. Overall the increase of $I3(6PR)$ remains intact, but there are strong spikes superimposed. The strongest peaks occur for the cases B63 and B70, where an additional uncited paper lets $I3(6PR)$ jump from 75 to 98 and from 80 to 111. Two and three steps, respectively, with additional citations are needed to compensate. Figure 1 ends with case B73, comprising 29 uncited, 15 singly cited, 9 doubly cited, and 7 triply cited papers, i.e., 60 papers with altogether 54 citations. Of course, this development from case B1 to case B73 is an extreme example, but it shows how frequently unexpected inconsistencies can occur.

As mentioned above, these problems do not appear when one averages over assigned weights for tied papers instead of assigning one weight for the averaged rank, or lowest rank, or highest rank of the tied papers. This can be seen in Figure 1 where the indicator $I3^{av}(6PR)$ based on the averaged weights is also plotted. It increases monotonously, each additional paper leads to a small increase, but these steps are unfortunately not all identical due to the above mentioned finite-size effects. If the number of papers does not change, then also $I3^{av}(6PR)$ remains constant as discussed above.

The respective behavior of the indicator $R(6,k)$ is presented in Figure 2a. Here the above mentioned jump of $R(6,k)$ from case B1 to B2, the subsequent decrease for the cases up to B15, the following jump for B16, and the subsequent erratic behavior with big fluctuations, strongest for the cases B63 and B70 can be easily identified.

The curve for the averaged indicator $R^{av}(6,k)$ in Figure 3a is much smoother, although there are still some fluctuations due to the discretization for the small finite numbers as discussed above. Accordingly, with increasing number of papers these fluctuations become smaller and the indicator converges towards the theoretical value 1.91, which is given by the proportions of the PR classes and their weights: 50 % * 1 + 25 % *2 + 15 % * 3 + 5 % * 4 + 4 % * 5 + 1 % * 6 = 1.91.

**Assigning the Papers of Example B to Four Individuals in a Simple Pattern**

The demonstrated fluctuations of the indicators are strange but they may not be considered to be a serious problem, because one might be tempted to allow such fluctuations for the description of the complete database which is used as a reference set. However, in the following I demonstrate that these fluctuations are strongly reflected in the behavior of the indicators for different subsets of the database. These subsets can be interpreted as the citation records of individual researchers contributing to the complete sample which might be interpreted as the accumulated data for the institution of these researchers being used as the reference set for the evaluation. It will be shown that this partitioning can lead not only to an erratic behavior of the indicators for the subsets but also to chaotic evolution of the ranking of the subsets, i.e. the individuals; and it will be shown that the proposed averaging of the weights will smooth the fluctuations as expected. For this purpose I return first to example B and Figure 2.

In Figure 2 it is also shown what can happen, if the papers are distributed among 4 scientists. This example is based on the evolution of the paper and citation counts in example B and

Figure 1. I have apportioned the papers in example B to 4 scientists H, L, M, N in a very simple way, attributing the 4 papers with 3 citations each to the highly cited individual H, the 4 papers with 2 citations each to the moderately cited scientist M, and the 7 papers with only one citation to the lowly cited researcher L. The papers and citations which have been added during the evolution of the example B from case B1 to case B73 in the previous section are all ascribed to the new colleague N. This is a rising star and it is no surprise that the indicator $R(6,k)$ in Figure 2a reflects this rise. However, again unreasonably strong fluctuations can be seen. Even more strange is the behavior of the resulting values of the individual indicators $R(6,k)$ for the 3 colleagues. Although they do not publish anything and also do not receive any citations, already the first paper of N leads to a strong increase of the indicators for H and M. The following smooth decrease for H, M, L and the respective increase for N up to the case B15 are reasonable. But then the indicators fluctuate so strongly that in many cases the rank order of H, M, L is mixed up, although they still do not publish and are also not cited anymore.

In Figure 2b the ranks are visualized and the erratic behavior becomes even clearer. These cases demonstrate that unreasonable changes in the ranking of the scientists occur, which means that not only the total value of $R(6,k)$ behaves in an inconsistent way, but that this leads also to inconsistencies in the ranking of individual contributors.

Figure 3a shows that the above proposed averaging of the weights leads to a much smoother behavior of the indicators, although some undesirable fluctuations remain due to the discretization effects for the small finite numbers of papers. But as Figure 3b demonstrates, the ranking of the scientists is now free from unexpected fluctuations: The newcomer N rises from the lowest rank to the top. The change of ranks between M and L can be explained by looking into the details of the calculation. In the beginning in cases B2 and B3 the 7 singly cited papers of scientist L each obtain a weight 1, while the 4 doubly cited papers of scientist M are weighted with a factor 2, leading to a slightly smaller contribution of 7 to $I3^{av}(6PR)$ for L in contrast to 8 for M. At the end of the evolution of example B the respective weights are 2 and 3, so that the contribution of 14 to $I3^{av}(6PR)$ for L is slightly larger than the contribution of 12 for M. This leads to an exchange of the ranks between M and L, compare cases B3 and B8 in Figure 3b, with tied ranks in between.[3] In principle, this behavior could also be called inconsistent, because the ranking changes for two scientists, whose publication numbers and citation records do not change. The fundamental reason for this behavior is the nonlinearity of the indicator. With the increasing number of uncited papers in the total dataset, the papers of L and M move to higher PR classes; due to the nonlinearity the relative weights change from the ratio 1:2 to 2:3. So this is a different problem from the above solved difficulty concerning the treatment of tied ranks. It will be discussed in the final section that this may not even be a problem but rather a desired property which enables a meaningful distinction between different fields where it should be possible that different reference sets lead to different relative weights.

---

[3] There is also a tie between M and L in the case B1, because the papers of M get a weight of 1.75 due to the averaging.

**Assigning the Papers to the Four Scientists in a More Complicated Pattern**

Finally I present a further example C, which is also based on the evolution of the paper and citation counts in example B. It is the purpose of example C to display an even more chaotic behavior and to demonstrate that this is also appropriately smoothed by the proposed procedure of averaging the weights. Moreover, this example will also be used in the next section for discussing the consistency of $R(6,k)$ when two scientists achieve the same performance improvement.

In this example C I have interrupted the advance of the rising star N after 10 publications and allowed the other scientists to add some uncited publications to their publication record. However, further citations are still attributed only to N. The actual sequence of additional papers and citations is indicated below the horizontal axis in Figure 4. The total number of papers and citations in each case does not change in comparison to the previous figures, only the attribution of the uncited papers changes. I have tried to select these changes in such a way, that the erratic behavior of the indices for the different scientists is maximized, but with the aim of ending the evolution with the same number of papers for everybody, i.e., 15 papers for each scientist.

The wild fluctuations of $R(6,k)$ in Figure 4 show that I have been rather successful. The respective rankings are visualized in Figure 4b. For example, a single additional paper for N in case C16 has moved scientist L from the lowest rank in C15 to the top in C16, but this windfall profit is immediately lost, when that publication of N receives a citation, see case C17. In case C19 it is at least an additional paper of L which has led to the same jump of L from last to first rank, but again an additional citation for N completely counterbalances this. In case C30 it is an additional publication of H which has led to the same rank jump for L, again counterbalanced by an additional citation of N in case C24. Likewise an additional publication of scientist M in case C49 has moved L from the last rank to the first place (tied with N). Here N is already rather advanced and keeps it top rank. Nevertheless, an additional citation again counterbalances the windfall profit of L. Similar changes, not always as dramatic as these can be found frequently. Particularly odd is case B41, where an additional paper of M does not change M's third position, but it pushes H from first to last place and L from last to second. In the next step B42, a citation to a paper of N leads to a drop of N and an advance of L, i.e., they exchange positions, and so do H and M. A further citation for N, case B43, does not change anything for N, but sends L down to last position and H up to top rank, i.e., restores the sequence of case B40.

Overall the fluctuations in the ranking for example C are even stronger than in example B. This problem can again be solved by the above proposed alternative calculation of the indicators averaging over the assigned weights for the tied papers. The results are shown in Figure 5a. The curves are not as smooth as in Figure 3a, but this cannot really be expected, because now the publication record of H, M, L is not anymore assumed to remain constant.

The respective rankings in Figure 5b show the drastic improvement in comparison to Figure 4b. The exchange of the second and third place in the ranking between M and L at the beginning of the evolution, compare cases C3 and C8, is the same as in example B as discussed above in connection with Figure 3b. But in contrast to example B, now the scientists M and L change ranks again between cases C49 and C50, so that in the end M is ranked better than L. The reason for this different final outcome is due to the different number of uncited papers which are attributed to M and L in order to achieve the final total numbers of 15 publications for everybody.

**Introducing Fractional Scoring**

Rousseau (2011, 2012) and Leydesdorff and Bornmann (2012) have discussed different scoring rules and their application for the indicator $I3$(6PR) without coming to a final solution, but rather fearing that they "may have opened a box of Pandora allowing for generating a parameter space of other possibilities" (Leydesdorff & Bornmann, 2012). Attributing averaged weights as proposed above appears to be a good solution. However, if one applies the above used counting rule one of the remaining problems is that the best paper never gets the highest score if the total number of papers is smaller than 100.

I would like to note that the ad-hoc suggestion of Leydesdorff and Bornmann (2011) to change the rounding by adding 0.9 to the paper count does not influence the calculation of the indicators in the present example. The authors had introduced this rounding to avoid undesirable effects for datasets that are smaller than 100. The above demonstrated inconsistency cannot be solved by this unusual rounding. Moreover, this rounding can also lead to other undesirable effects like putting already the 110[th] of 111 papers into the top-1% class (assuming that 109 papers have less citations, yielding a count of 109.9, which is 99.01 % of 111 papers) which thus would comprise two of 111 papers, i.e. nearly 2%. So it has more than the claimed "marginal effects for numbers in the set larger than 100" (Leydesdorff & Bornmann, 2012).

A slightly different scoring rule in agreement with Rousseau (2012) always attributes the highest score to the most cited paper. For this purpose the papers are sorted by increasing number[4] of citations and given the rank $r$ without caring about the sequence of tied papers as above. Then the percentage $r * 100 / n_{tot}$ is calculated and compared with the boundaries of the six PR classes. To be specific, papers with thus determined percentage less or equal than 50.0 are included in class 1, and so on. Note that in contrast to the above utilized scoring rule now the paper in question is included in the counting and correspondingly the percentage boundary is also included in the comparison (therefore the rule is now less or equal instead of less as above). With regard to example A, this change of the counting rule does not influence the determination of $I3^{av}$(6PR) except for the top-ranked paper which is now attributed a weight of 6 instead of 5. This leads to an increase of $I3^{av}$(6PR) to 77, correspondingly $R^{av}(6,k)$ = 1.925 for all cases in Table 3. For examples B and C, more changes occur, because often papers close to but not exactly at the borders between the PR classes are given a different weight.

However, a closer inspection reveals that this rule is still not perfect, because the final result deviates from the theoretical value 1.91. The deviation can be traced to the discretization. In the present example A with 40 papers every paper amounts to 1/40 = 2.5 % of the total number. Accordingly the top ranked paper contributes the highest weight $w$ = 6 with a proportion of 2.5 % instead of 1 % in the theoretical distribution. On the other hand using the scoring rule from Leydesdorff et al. (2011) as in Tables 1 - 3 the top 2 papers would contribute the weight $w$ = 5 with a proportion of 2 * 2.5 % = 5 % instead of 4 % (plus 1 % of $w$ = 6) as in the theoretical distribution.

The solution of this problem is a new scoring rule which is independent of the specific way in which papers at the boundaries of the six or hundred PR classes are treated. In this new scoring rule only the 1 % fraction of the top paper is given the weight $w$ = 6 and the remaining fraction 1.5 % is given the weight $w = 5$. This would amount to a contribution[5] of 1 % * 6 + 1.5 % * 5 = 0.135 to $R^{fr}(6,k)$ where I have introduced the superscript "fr" to indicate the use of the fractional scoring rule. Concerning the calculation of $I3$(6PR) this fractionalization of the 2.5 % of the total number into 1 % of the top class and 1.5 % of the second highest class means that two fifths of the paper get the weight $w = 6$ and three fifths get the weight $w = 5$, so that $I3^{fr}$(6PR) = 0.4 * 6 + 0.6 * 5 = 5.4. This value should not be rounded to 5, but taken as it is. For completeness I note that in the case of tied papers as mentioned above, after assigning the fractional weights to the tied papers one has to average the weights of the tied papers and reassign the average weight to these tied papers.

---

[4] It should be noted that percentiles are more often determined after ranking the papers by decreasing number of citations.

But this fractional scoring makes the determination of the weights rather complicated in the general case, because for a different total number $n_{tot}$ such a fractional attribution of weights would have to be applied not only for the top-ranked paper, but at all or nearly all borders between the different PR classes. In the present example A all borders except for the 99 % boundary coincide exactly with one of the $r/n_{tot}$ values. Usually this is not the case and thus a fractional attribution of the weights would be always necessary.[6] Although it is more significant for datasets with a small number of papers, the scoring rule should be applied also for large datasets, because in the general case it can make a difference for all manuscripts which are closest to a border between different rank classes. In conclusion I propose to always apply fractional scoring, because together with the suggested averaging of the fractional weights for the tied papers it solves all discussed problems and thus closes the above mentioned box of Pandora.

**Concluding Remarks**

---

[5] In contrast in Tables 1-3 the top paper got a weight of 5 and contributed 5/40 = 0.125 to $R(6,k)$ and $R^{av}(6,k)$, except in case A8 where it achieved only the weight 4 and thus a contribution of 0.1 to $R(6,k)$; for $R^{av}(6,k)$ the top paper would get the weight 5 also in case A8 (before averaging over the tied papers). In all these cases, the results are smaller than for the fractional scoring. On the other hand, the above mentioned scoring rule from Rousseau (2012) would always assign a weight of 6 to the top paper and would thus always lead to a contribution of 6/40 = 0.15 which is above the fractional scoring value.

[6] For very small numbers $n_{tot} < 20$ as occurring in the examples B and C, the top-ranked paper would even be spread over 3 PR classes and would have to be attributed 3 different weights fractionally from the top 3 PR classes. For example, in the case $n_{tot} = 16$ every paper amounts to 1/16 = 6.25 % and the top paper receives 1.25 % of weight 4 from the fourth PR class, the full 4 % of weight 5 from the fifth PR class and the full 1 % of weight 6 from the top class, altogether a contribution of 1.25 % * 4 + 4 % * 5 + 1 % *6 = 0.31 to $R^{fr}(6,k)$.

Waltman and van Eck (2012) have discussed the consistency of performance indicators from a theoretical point of view. Formal aspects can be found in the literature cited by Waltman and van Eck (2012). In that paper the theoretical argument is presented completely in intuitive terms, and in particular the authors demand that *"If two scientists achieve the same absolute performance improvement, their ranking relative to each other should remain unchanged."* Although I have not constructed my example C with regard to this definition of consistency, a closer inspection of Figure 4 shows several cases in which this condition is violated. Due the specific evolution of the example cases, there are only 15 instances in which two of the scientists achieve the same performance improvement, namely the addition of an uncited paper. Specifically, comparing cases C17 and C19, H and L both get one more uncited paper, but exchange their relative ranking. The same happens for them between cases C43 and C45. Similarly, between cases C39 and C41 the relative ranking of M and L is changed although both of them achieve the same performance improvement of one additional uncited paper. And while in case C54 there is a tie between N and H on ranks 2 and 3, attributing one more uncited paper to each of them leads to ranks 1 for N and 4 for H in case C56. Thus in 4 out of 15 instances the indicator behaves in an inconsistent way in example C.[7]

An inspection of Figure 5b shows that the proposed averaging of the weights leads to a consistent behavior: in all mentioned cases the relative ranking remains unchanged.

But in view of the strong fluctuations of the indicators and the rankings in Figure 2 I propose a further property of strict independence,[8] which a performance indicator should aim at:

> *"The ranking of two scientists relative to each other should remain unchanged, if a third scientist achieves a performance improvement."*

In example B, in which the performance improvement is solely attributed to scientist N, this condition is violated frequently as all the changes in the rankings among H, M, and L in Figure 2b visualize. Again averaging the weights, the newly proposed indicator behaves in a much more consistent way, as demonstrated by the rankings in Figure 3b which do not fluctuate. Besides the reasonable advance of N from the bottom to the top rank there is only one change in the ranking, between M and L, which has already been discussed above and explained by the nonlinearity of the indicator. Thus, with this exception, the alternative calculation of the indicators based on averaging over the assigned weights fulfills the newly proposed property of strict independence.

---

[7] In the evolution of example B only the performance of scientist N is changed, so that there cannot be any instance in which two scientists achieve the same performance improvement.
[8] Following Rousseau (2012) I use the term independence rather than consistency. In the present case this terminology reflects the situation better, because it refers to the requirement that the ranking of two scientists should be independent from the performance improvement of a third scientist.

It should be noted that this property is trivially fulfilled by unnormalized indices like the h-index which are only a function of the publications of an individual author and the citations to these publications, not using any other information. For normalized indicators like $I3(6PR)$ and $R(6,k)$ one needs a reference set which in the present analysis has been taken to be the accumulated citation distribution of the 4 scientists. Different reference sets allow the comparison between different fields and it is therefore desirable that completely different reference sets can lead to different rankings of groups of scientists with identical citation records, but in different fields. In the present examples, the additional publications and citations change the reference set and thus the various cases could be interpreted as describing a variety of fields. Then it might be expected that the proposed criterion of strict independence is not fulfilled. However, it is odd that a series of small changes violates the property as strongly and as often as shown in figures 2b and 4b. In my view it is much more reasonable that the ranking changes much more smoothly like in figures 3b and 5b.

The fluctuations in the presented examples occur due to tied papers. I have intentionally constructed examples with a relatively small number of papers and with rather small citation frequencies, in order to enhance the inconsistencies. In a more realistic sample, there would be fewer ties and thus the jumps and drops of the indicators would happen less frequently. Moreover, usually the reference set would be much larger and at least comprise the citation records of all scientists of an institution, i.e. of much more than 4 people. Then small changes in the reference set would not have such a big influence as in the presented examples. Analyzing and comparing complete institutions or journals, or even countries the data base would probably be so large that usually such problems would not occur at all. And if one takes all publications in a given field as the reference set which is used for the normalization and measures the citation records of individual scientists or institutes against this reference set, then it is extremely unlikely that the small changes have any influence on the ranking at all. Nevertheless, as a matter of principle these indicators should be defined in a way to avoid inconsistencies. As has been shown above, by treating the tied papers in a different way the erratic fluctuations can be drastically reduced. Changes in the ranking can and should not be completely avoided, because the nonlinearity of the weights for the different PR classes may still lead to a different ranking, what is reflecting a difference in the reference set and thus desired when comparing different fields.

**Acknowledgement**

I am grateful to two reviewers for very useful suggestions. Especially I thank L. Waltman for the suggestion about the alternative assignment of weights in fractional parts.

TABLE 1. Determination of the indicator $I3(6PR)$ for case A1 of example A comprising $n(c)$ papers with $c$ citations each. There are no papers with $c = 2, 4, 6, 8$, or more citations, so that respective columns are excluded.

| example A | citation frequency $c$ | | | | | total |
|---|---|---|---|---|---|---|
| case A1 | 0 | 1 | 3 | 5 | 7 | |
| number $n(c)$ of papers | 20 | 10 | 6 | 2 | 2 | 40 |
| citations $c*n(c)$ | 0 | 10 | 18 | 10 | 14 | 52 |
| number $n_<$ of papers with less citations | 0 | 20 | 30 | 36 | 38 | |
| percentage $n_<*100/n_{tot}$ | 0 | 50 | 75 | 90 | 95 | |
| rank class = weight $w$ | 1 | 2 | 3 | 4 | 5 | |
| weighted number of papers $w*n(c)$ | 20 | 20 | 18 | 8 | 10 | 76 |

TABLE 2. Same as Table 1, but for case A2 in which one uncited paper from A1 received its first citation.

| example A | citation frequency $c$ | | | | | total |
|---|---|---|---|---|---|---|
| case A2 | 0 | 1 | 3 | 5 | 7 | |
| number $n(c)$ of papers | 19 | 11 | 6 | 2 | 2 | 40 |
| citations $c*n(c)$ | 0 | 11 | 18 | 10 | 14 | 53 |
| number $n_<$ of papers with less citations | 0 | 19 | 30 | 36 | 38 | |
| percentage $n_<*100/n_{tot}$ | 0 | 47.5 | 75 | 90 | 95 | |
| rank class = weight $w$ | 1 | 1 | 3 | 4 | 5 | |
| weighted number of papers $w*n(c)$ | 19 | 11 | 18 | 8 | 10 | 66 |

TABLE 3. Development of the indicators $I3(6PR)$ and $R(6,k)$, if the uncited paper from case A1 receives more and more citations, as specified step by step by the number of papers for the different citation frequencies.

| example A | number of papers with citation frequency | | | | | | | | | total number of papers | total number of citations | $I3(6PR)$ | $R(6,k)$ |
|---|---|---|---|---|---|---|---|---|---|---|---|---|---|
| case | 0 | 1 | 2 | 3 | 4 | 5 | 6 | 7 | 8 | | | | |
| A1 | 20 | 10 | 0 | 6 | 0 | 2 | 0 | 2 | 0 | 40 | 52 | 76 | 1.90 |
| A2 | 19 | 11 | 0 | 6 | 0 | 2 | 0 | 2 | 0 | 40 | 53 | 66 | 1.65 |
| A3 | 19 | 10 | 1 | 6 | 0 | 2 | 0 | 2 | 0 | 40 | 54 | 67 | 1.68 |
| A4 | 19 | 10 | 0 | 7 | 0 | 2 | 0 | 2 | 0 | 40 | 55 | 61 | 1.53 |
| A5 | 19 | 10 | 0 | 6 | 1 | 2 | 0 | 2 | 0 | 40 | 56 | 62 | 1.55 |
| A6 | 19 | 10 | 0 | 6 | 0 | 3 | 0 | 2 | 0 | 40 | 57 | 60 | 1.50 |
| A7 | 19 | 10 | 0 | 6 | 0 | 2 | 1 | 2 | 0 | 40 | 58 | 61 | 1.53 |
| A8 | 19 | 10 | 0 | 6 | 0 | 2 | 0 | 3 | 0 | 40 | 59 | 59 | 1.48 |
| A9 | 19 | 10 | 0 | 6 | 0 | 2 | 0 | 2 | 1 | 40 | 60 | 60 | 1.50 |

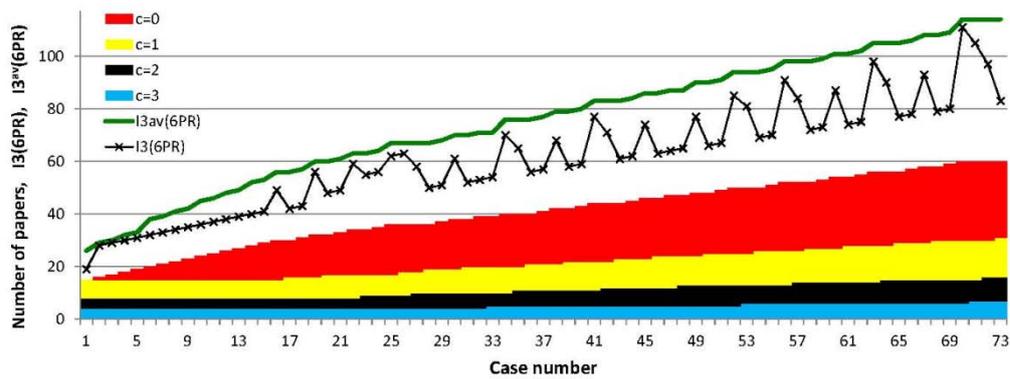

FIG. 1: Evolution of indicators $I3$(6PR) (thin black line with crosses) and $I3^{av}$(6PR) (thick green line) with increasing number of papers and citations for example B, where the number of papers with 0, 1, 2, and 3 citations is given by the histogram (from top to bottom).

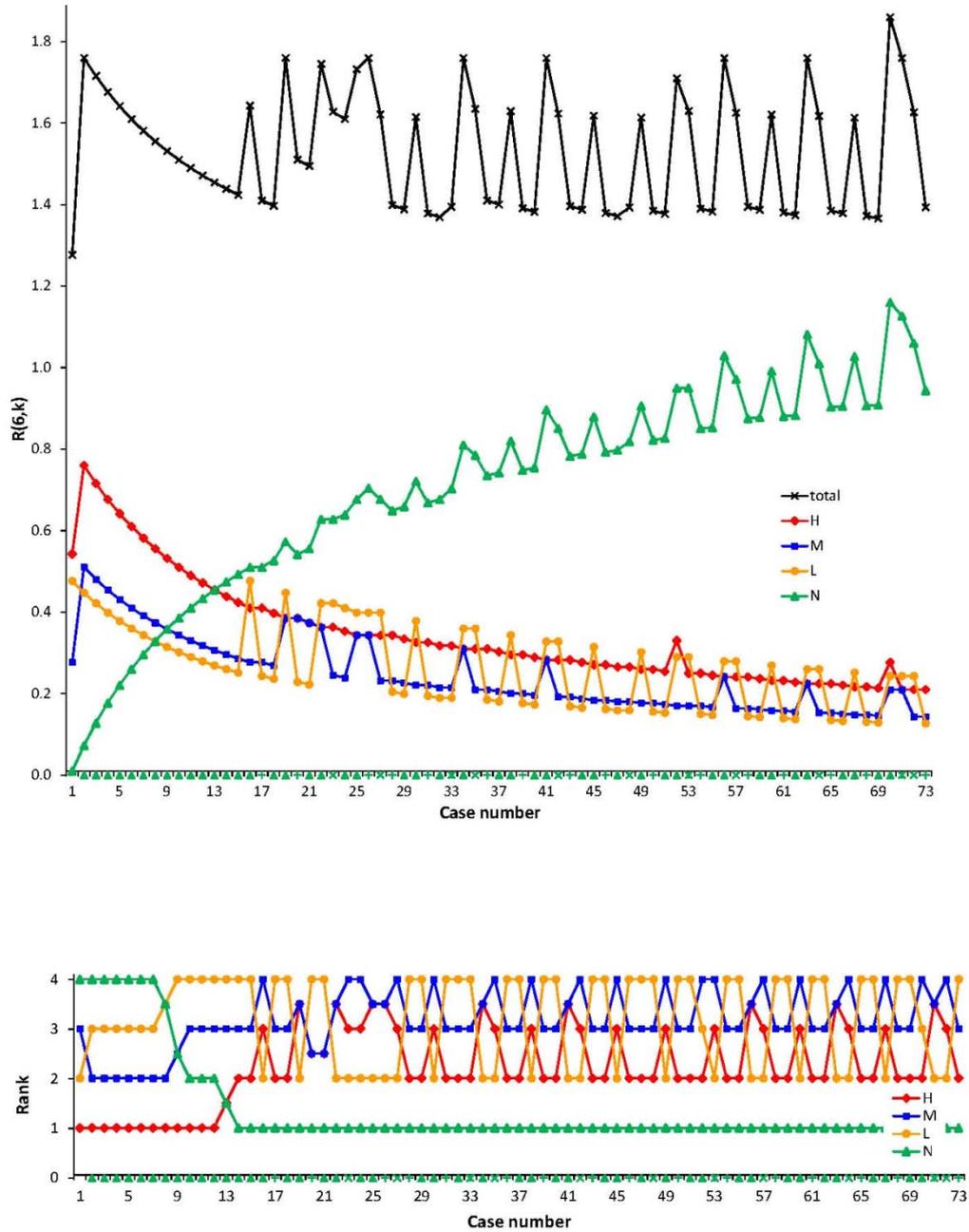

FIG. 2. (a) Evolution of the indicator $R(6,k)$ (thin black line with crosses) for example B and contributions of the 4 scientists H, L, M, N as described in the text. (b) Corresponding changes in the ranking of the 4 scientists, with the average value given for tied ranks. The symbols below the horizontal axis denote the changes in the dataset: a triangle indicates an additional paper for scientist N, a plus, cross, star means an additional citation to a previously uncited, singly cited, doubly cited paper of scientist N, respectively. By construction of the example, exactly one of these changes occurs between subsequent cases.

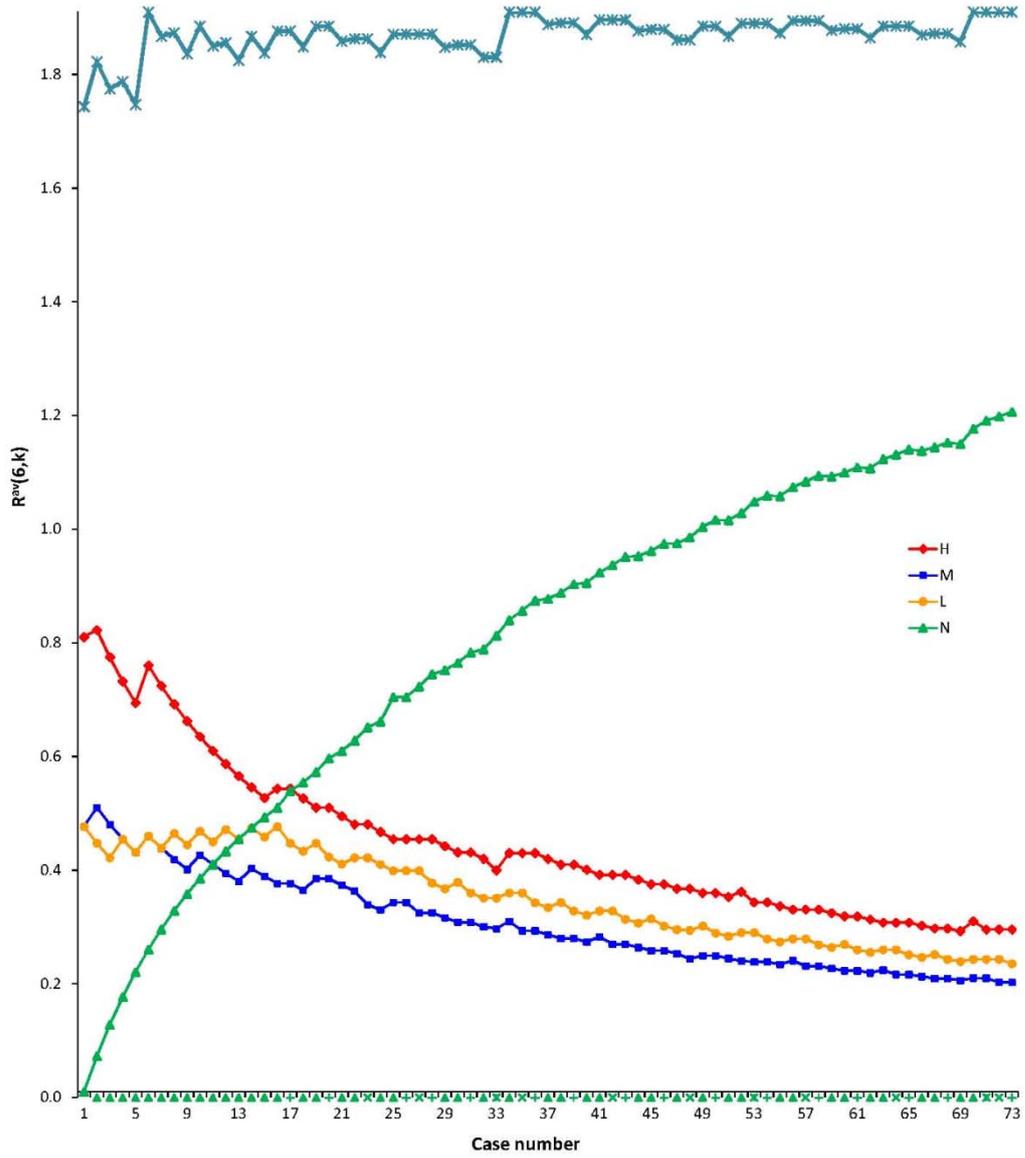

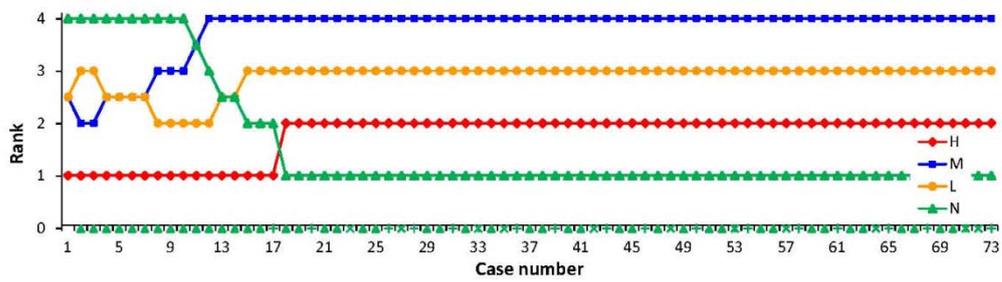

FIG. 3. Same as Figure 2, but for the indicator $R^{av}(6,k)$, i.e., for averaged weights.

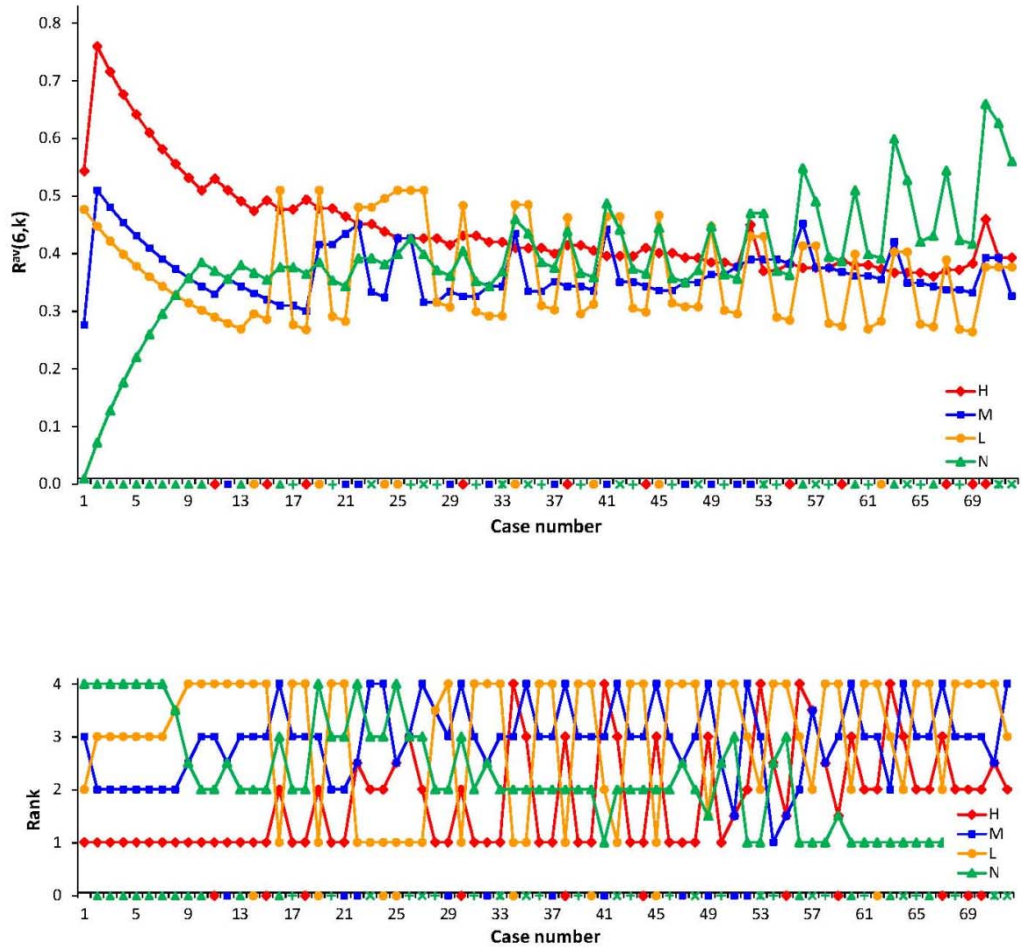

FIG. 4. Same as Figure 2, but for example C. The total number of papers and citations in each case is the same as in example B, so that the summed indicator $R(6,k)$ is not shown again. The symbols below the axis have the same meaning as in Figure 2, additionally a red diamond, a blue square, and an orange circle indicate an additional paper for scientist H, M, L, respectively. By construction of the example, exactly one of these changes occurs between subsequent cases.

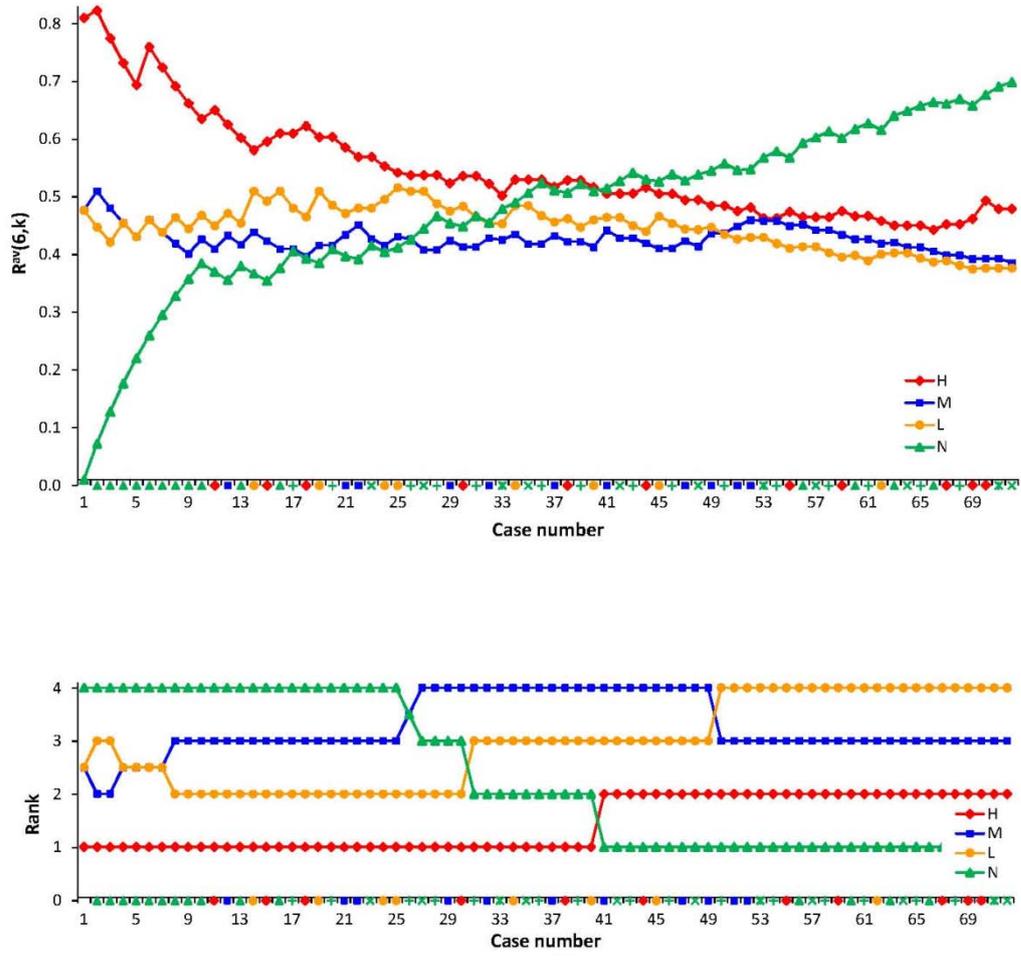

FIG. 5. Same as Figure 4, but for the indicator $R^{av}(6,k)$, i.e., for averaged weights.